\documentclass[aps,showpacs, nofootinbib]{revtex4}

\usepackage{graphicx}

\newcommand{\be}{\begin{equation}}
\newcommand{\ee}{\end{equation}}
\newcommand{\ben}{\begin{eqnarray}}
\newcommand{\een}{\end{eqnarray}}

\newcommand{\la}{{\lambda}}

\newcommand{\na}{\nabla}

\newcommand{\tpsi}{\tilde \psi}

\newcommand{\tn}{{\tilde n}}


\begin{document}

\title{Evolution of a Massive Scalar Fields in the Spacetime of a Tense Brane Black Hole} 

\author{Marek Rogatko and Agnieszka Szyp\l owska}
\affiliation{Institute of Physics \protect \\
Maria Curie-Sklodowska University \protect \\
20-031 Lublin, pl.~Marii Curie-Sklodowskiej 1, Poland \protect \\
rogat@kft.umcs.lublin.pl\protect \\
marek.rogatko@poczta.umcs.lublin.pl}

\date{\today}

\begin{abstract}
In the spacetime of a $d$-dimensional static tense brane black hole we elaborate
the mechanism by which massive scalar fields decay. The metric of a six-dimensional 
black hole pierced by a topological defect is especially interesting. 
It corresponds to a black hole residing on a tensional 3-brane embedded in a 
six-dimensional spacetime, and this solution has gained importance due to the planned accelerator experiments.
It happened that the intermediate asymptotic behaviour of the fields in question was
determined by an oscillatory inverse power-law. We confirm our investigations by numerical calculations
for five- and six-dimensional cases. It turned out that the greater the brane tension is, 
the faster massive scalar fields decay in the considered spacetimes.

\end{abstract}

\maketitle
\section{Introduction}
\label{intro}
In recent years the subject of the late-time behaviour of various fields in the spacetime of a collapsing
body has acquired a great deal of attention due to the fact that
regardless of details of the collapse or the structure and properties
of the collapsing body the resultant black hole may be described
by a few parameters such as mass, charge and angular momentum, {\it black holes have no hair}.
\par
In \cite{pri72} it has been pointed out 
that the late-time behavior is dominated by the
factor $t^{-(2l+3)}$, for each multipole moment $l$, while
in \cite{gun94} that the decay-rate along null infinity
and along the future event horizon is governed by the power laws
$u^{-(l+2)}$ and $v^{-(l+3)}$, where $u$ and $v$ were the outgoing
Eddington-Finkelstein (ED) and ingoing ED coordinates.  
The scalar perturbations in the  Reissner-Nordtsr\"om (RN) spacetime
were treated in \cite{bic72}. It happened that a charged hair 
decays slower than a neutral one
\cite{pir1}-\cite{pir3}.  Burko~\cite{bur97} reported 
the late-time tails in gravitational
collapse of a self-interacting (SI) fields in the background of
Schwarzschild solution. On the other hand, RN
solution at intermediate late-time was considered in
\cite{hod98}. The nearly extreme case of the RN spacetime
was elaborated analytically in article \cite{ja}.It was concluded that
the inverse power law behavior of the dominant asymptotic tail is of the
form $t^{-5/6} \sin (m t)$. It turned out \cite{ja1} that 
the oscillatory tail of scalar field in Schwarzschild spacetime
has the decay rate of $t^{-5/6}$ at asymptotically late time.
The power-law tails in the evolution of a
charged massless scalar field around a fixed background of dilaton
black hole was studied in \cite{mod01a}, while the case of a
self-interacting scalar field was elaborated in
 \cite{mod01b}.  The
analytical proof of the above mentioned behaviour of massive scalar
hair in the background of a dilaton black hole in the theory with
arbitrary coupling constant between $U(1)$ gauge field and dilaton
field was given in \cite{rog07}.
\noindent
It was revealed \cite{jin04} that the asymptotic behaviour
of massive Dirac fields in the background of Schwarzschild black hole
was independent of the multipole number of the wave mode and of the mass of
the Dirac field. Their decay was slower comparing to the decay of a massive scalar field.
The behaviour of Dirac fields in the spacetime of RN black hole was treated in \cite{jin05}
while the case of a stationary axisymmetric black hole
background was studied numerically in \cite{bur04}.
In \cite{xhe06} both the intermediate late-time tail and the
asymptotic behaviour of the charged massive Dirac fields in the
background of a Kerr-Newmann black hole was under consideration and it was
demonstrated that the intermediate late-time behaviour of the fields in question
was dominated by an inverse power-law decaying tail
without any oscillation.
\noindent
Massive vector field obeying the Proca
equation of motion in the background of Schwarzschild black hole was
studied in \cite{kon06} where it was revealed that at intermediate late
times, three functions characterizing the field have different decay
law depending on the multipole number $l$.  On the contrary, the
late-time behaviour is independent on $l$, i.e., the late-time decay
law is proportional to $t^{-5/6}\sin(mt)$.  In \cite{gibrog08} the
analytical studies concerning the intermediate and late-time decay
pattern of massive Dirac hair on the dilaton black hole were
conducted.  Dilaton black hole constitutes a static spherically
symmetric solution of the theory being the low-energy limit of the
string theory with arbitrary coupling constant $\alpha$.
In \cite{mod08} it was assumed that the massive Dirac hair propagated in
a static spherically symmetric spacetime with asymptotically flat metric. 
The intermediate late time behaviour was found numerically and the final
decay pattern at very late times was calculated analytically. The metric 
in question, was characterized by the ADM mass and some other parameters of the background
fields. The intermediate and
late-time decay of massive scalar hair on static brane black holes was
elaborated in \cite{rog07br}, while the decay of massive Dirac hair in the spacetime of black hole in question
was studied in \cite{gibrog08br}.
\par
For $n$-dimensional static black holes, the {\it no-hair}
theorem is quite well established \cite{unn}. One should also 
mention about
some attempts to establish uniqueness theorem for another kind of black objects, black rings. 
The uniqueness theorem for five-dimensional stationary the so-called Pomeransky-Sen'kov black ring was
found in \cite{mor08} while black ring $\sigma$-model uniqueness theorem was given in \cite{rog08}.
On the other hand,
the mechanism of
decaying black hole hair in higher dimensional static black hole case
concerning the evolution of massless scalar field in the
$n$-dimensional Schwarzshild spacetime was determined in
\cite{car03}.  The late-time tails of massive scalar fields in the
spacetime of $n$-dimensional static charged black hole was elaborated
in \cite{mod05} and it was found that the intermediate asymptotic
behaviour of massive scalar field had the form $t^{-(l + n/2 -
1/2)}$. This pattern of decay was checked numerically for $n = 5, 6$.
Quasi-normal modes for massless Dirac field in the background of
$n$-dimensional Schwarzschild black hole were studied in
\cite{cho07a}.
\par
Due to the fact that forthcoming Large Hadron Collider (LHC) may open the way for producing 
microscopic black holes which will quickly decay through the Hawking radiation,
interest in tensional brane black holes rapidly grows. In  \cite{kal06}
the exact metric describing a black hole localized on a codimensional-2 brane was constructed.
It turned out that the line element $d^2 \Omega_{d - 2}$ on the unit $(d - 2)$-dimensional sphere
is replaced by the metric on a unit sphere but with the wedge removed from the polar coordinates, i.e.,
one gets some kind of a {\it topological defect} piercing the black hole in question.
The finite brane tension modifies the standard results achieved in the case of a brane black hole
with negligible tension. 

This discovery bears directly on the question of the underlying physics. Namely,
for the first time it was observed in the semi-classical description
of the black hole decay process \cite{dai07}.
In \cite{kil07} the metric of codimensional brane black hole was generalized to the rotating case, while article
\cite{cho07b} was devoted to massless fermion
excitations on a tensional three-brane embedded in six-dimensional
spacetime. 
Various aspects of the tense brane black hole physics were studied.
Among all,
quasi-normal modes, bulk scalar emission, grey-body factors 
and the behaviour of scalar perturbations in the background of a black hole localized
on a tensional three-brane in a world with two large extra dimensions were elaborated
in \cite{bin07}-\cite{che07b}.
Having all in mind it will not be amiss to pay more attention to the decay process of massive scalar
hair in the background of a tense brane black hole.
\par
In this paper our investigations will be devoted to studies
of $d$-dimensional tensional brane black hole. In the case of $d = 5,~6$
we provide numerical analysis of the intermediate late-time behaviour of massive scalar fields
in the black hole backgrounds in question. 
\par
The remainder of the paper is as follows.
In Sect. \ref{sec2} we shall discuss the massive scalar field behaviour in the spacetime
of $d$-dimensional static spherically symmetric tense brane black hole.
In Sect. \ref{numer} we treat numerically the problem of five and six-dimensional black hole in question.
We conclude our investigations in Sect. \ref{con}.

\section{Massive scalar fields in a tense brane black hole background}
\label{sec2}
Recent progress in string theory and brane world models led us to an intense interest
in higher-dimensional black holes. In model with large extra dimensions black hole size is much
smaller than the effective size of the extra dimensions. Simply additional dimensions are considered as 
having infinite extent. Models with large extra dimensions acquire much more attention also 
because of 
an interesting possibility of lowering the fundamental scale of gravity down to order of TeV. 
It can be argued that future experiments, which involve high-energy
particles collisions at future colliders, hold great promise for illuminating the nature of mini black holes.
Especially, forthcoming LHC brings the possibility of creating microscopic black holes which  
will disappear quickly after creation with the emission of the Hawking radiation. Most examinations of extra-dimensions
black holes and their physics were devoted to the zero brane tension case. But in principle
finite brane tension may modify the physics of a tensional brane black hole. 
The nonzero tension on the brane can curve the brane as well as the bulk.
Therefore more attention should be directed to the studies of the afore mentioned black objects.
A tense brane black hole is locally a higher-dimensional Schwarzschild solution \cite{kal06}
threaded by a tensional brane. This caused that a deficit angle appeared in the $(d - 2)$-dimensional
unit sphere line element. In what follows, our considerations will be devoted to studies of such kind of 
black objects pierced by a codimension-2 brane.
\par
In our considerations we take into account the following equation of motion
for massive scalar field:
\be
\na^{i}\na_{i}\tpsi - m^2 \tpsi = 0,
\ee
where $m$ is the mass of the field. One should study its behaviour 
in the background of
static spherically symmetric 
$d$-dimensional black hole. The line element of such a black hole is subject to the relation
\be
ds^2 = - f^2 (r)dt^2 + \frac{dr^2}{f^2 (r)} + r^2 d\Omega^2_{d-2},
\ee
where $f^2 (r) = 1 - \bigg( r_{0} / r \bigg)^{d -3}$, $r_{0}$ is the radius of the black hole event horizon,
while $d\Omega^2_{d-2}$ is a line element
on $S^{d-2}$ sphere provided by the relation
\be
d\Omega^2_{d-2} = d\theta^2 + \sum_{i = 2}^{d - 3} \prod_{j =1 }^{i -1} \sin^2 \phi_{j} d\phi_{j}.
\ee
In our considerations we take into account $S^{d-2}$ sphere threaded by a codimension-2 brane, so
the range of one of the angles, let us say $\phi_{i}$, will be $ 0 \le \phi_{i} \le 2 \pi B$.
\par
Consequently, in the case of $d$-dimensions 
we resolve the scalar field function as follows:
\be
\tpsi = \sum_{l,n} {1 \over r^{{d-2 \over 2}}} \psi_{n}^{l}(t, r) 
Y_{l}^{n}(\theta, \phi),
\label{sp}
\ee
where $Y_{l}^{n}$ are scalar spherical harmonics on the unit $(d-2)$-sphere. 
For the case when $(d-2)$ sphere is pierced by a topological defect, one can introduce
{\it hyper-spherical} coordinates in the form:
$0 < r < \infty$, $0 \le \theta \le \pi$, $-\pi \le \phi_{1} \le \pi,~\dots$, \mbox{$-\pi B \le \phi_{d-3} \le \pi B$.}
Having in mind results of \cite{coe02},
one may introduce $k = \sum_{i = 0}^{d - 4} n_{i}$, where $n_{i}$ are the integers separation constants
and now the multipole number $l$ has the form 
\be
l = k + \tn,
\ee
where $\tn = {n \over B}$ expresses the existence
of a topological defect. Further on,
it is convenient to make the change of variables. Namely,
let us define the tortoise coordinates $y$
\be
dy = {dr \over 
\bigg(1 - \bigg( {r_{0} \over r} \bigg)^{d -3} \bigg)}.
\label{coo}
\ee
Using Eq.(\ref{sp}) and relation (\ref{coo})
one has the following equations for each multipole moment:
\be
\psi_{,tt} - \psi_{,yy} + V \psi = 0,
\label{mo}
\ee
where the 
form of the strictly radial potential $V(r)$ yields
\be
V = f^2(r) \bigg[ \bigg( {d - 2 \over 2} \bigg)
{1 \over r^{{d-2 \over 2}}} {d \over dr} 
\bigg( r^{{d\over 2} - 2}~ f^2(r) \bigg) + {\bigg( k + \mid \tn \mid \bigg)~\bigg( k + d - 3 + \mid \tn \mid \bigg)
\over r^2} + m^2
\bigg].
\ee
The square of the Laplace operator on $S^{d - 2}$
threaded by a topological defect is given by
\be
\la^2 = \bigg( k + \mid \tn \mid \bigg)~\bigg( k + d - 3 + \mid \tn \mid \bigg).
\ee
One can introduce an auxiliary variable $\xi$ which satisfies $\xi = f(r) \psi$. Next,
one assumes that the observer and the initial data are in the region where the following condition is satisfied
$r \ll {r_{0} \over (r_{0} m)^2}$ and moreover we consider
the intermediate asymptotic behaviour of massive scalar field, i.e.,
$r \ll t \ll {r_{0} \over (r_{0} m)^2}$.
If we consider further, the explicit dependence on time in function
$\psi$ of the form $e^{- i omega t}$, where $\omega$ is frequency, then
we achieve
\be
{d^2 \xi \over d r^2} + \bigg[ \omega^2 - m^2
- {\nu (\nu + 1) \over r^2} \bigg] \xi = 0,
\label{wave}
\ee
where $\nu = \bigg( k + \mid \tn \mid - 2 + {d \over 2} \bigg)$. Solution of Eq.(\ref{wave})
can be found by virtue of the method used
in \cite{hod98d}. It happened  that the asymptotic intermediate behaviour of scalar 
field at fixed radius implies
\be
\psi \sim t^{- (k + \mid \tn \mid + {d\over 2} - {1 \over 2})},
\label{osc1}
\ee
while the intermediate behaviour of massive scalar fields at the future black hole horizon $H_{+}$ is dominated by an
oscillatory power law tails of the form as follows:
\be
\psi \sim v^{- (k + \mid \tn \mid + {d\over 2} - {1 \over 2})} \sin (mv),
\label{osc2}
\ee
where $v$ is is advanced null coordinate.\\
In what follows we shall treat the problem of five and six-dimensional static spherically symmetric
black hole threaded by a topological defect.
The six-dimensional case is of a special interest. In spherical coordinates its
line element implies
{\setlength\arraycolsep{2pt}
\begin{eqnarray}
ds^2 & = & - \left[ 1 - \left( {r_{0} \over r} \right)^3 \right] dt^2 +
{dr^2 \over \left[ 1 - \left( {r_{0} \over r} \right)^3 \right]} +{} \nonumber \\
 & & {}+r^2 (d\theta^2 + \sin^2 \theta [d\phi^2 +
\sin^2 \phi (d\chi^2 + B^2 \sin^2 \chi d^2 \psi)]).
\end{eqnarray}}
Now, the parameter $B = 1 - {\la \over 2 \pi M_{*}}$ measures the deficit angle about an 
axis which is parallel to the three-brane with finite tension $\la$. $M_{*}$ is the fundamental
mass scale of six-dimensional gravity. The black hole horizon is at the distance of 
$r_{0} = r_{s}/B^{1/3}$, where $r_{s}$ is the six-dimensional Schwarzschild radius
given in terms of the ADM mass by the relation
\be
r_{s} = \left( 
{1 \over 4 \pi^2 } \right)^{1 \over 3}
{1 \over M_{*}} \left( {M_{BH} \over  M_{*}} \right)^{1 \over 3}.
\ee
One can remark that the main effect of the brane tension is to change the relation 
between the black hole mass and the event horizon radius.
\par
In the next section we shall check our predictions numerically confining our attention
to the case of $d = 5$ and $d = 6$.

\section{Numerical results}
\label{numer}
Due to the difficulties in solving differential Eqs. of motion in terms of adequate special 
functions for arbitrary spacetime dimension, we treat them numerically. 
By the method presented in \cite{gun94,mod05} we shall find the solution of the equation of motion transformed into
retarded and future null coordinates
$(u, v)$. Consequently, we achieve the following:
\be
4 \psi_{,uv} + V \psi = 0.
\label{num}
\ee
Equation (\ref{num}) will be solved on an uniformly spaced grid using the explicit difference
scheme. As was pointed out previously, the late time evolution of a
massive scalar field is independent of the form of the initial data. In order to perform
our calculations we start with a Gaussian pulse of the form 
\be
\psi(u=0,v) = A \exp \left ( - {(v-v_0)^2 \over \sigma^2} \right )
\ee
Because of the linearity of the relation (\ref{mo}) one has freedom in choosing
the value of the amplitude $A$. For our purpose we fix it as $A = 1$.
The rest of the initial field profile parameters will be taken as $v_{0} = 50$ and
$\sigma = 2$.
\par
\begin{figure}
\begin{center}
\includegraphics[width=0.85\textwidth]{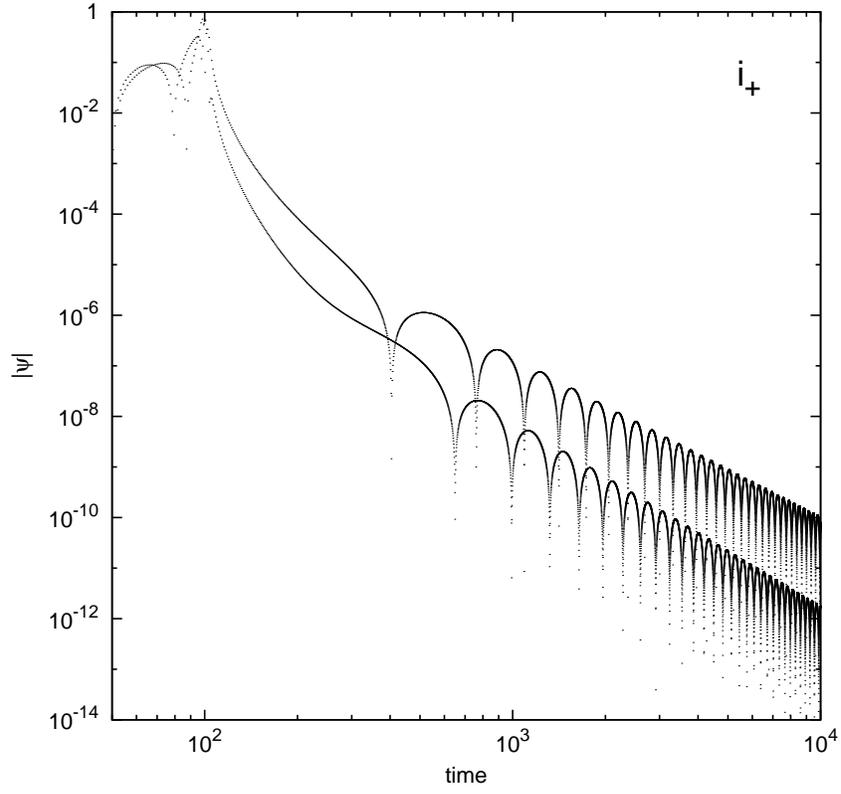}
\end{center}
\caption{Evolution of the scalar field $\mid \psi \mid$ on the future timelike infinity
$i_{+}$ as a function of $t$ for different spacetime dimensions $d = 5,~6$. Calculation parameters:
$m = 0.01,~k = 0,~n = 1,~ B = 0.9$. The power-law exponents are:
$- 3.14$ and $- 3.64$ for $d = 5,~6$, respectively. The period of oscillation is $T = {\pi \over m}
\simeq 314.5 \pm 0.5$ for both curves.}
\label{fig1}
\end{figure}

We begin our numerical studies with the evolution of scalar field $\psi$ on a future timelike infinity $i_{+}$.
In the considerations we approximate this case by the field at fixed radius $y = 50$. The numerical
results for $k = 0,~n = 1,~B = 0.9$ and mass of the field $m = 0.01$ for different spacetime dimensions are depicted in Fig. \ref{fig1}.
One should remark that the initial evolution of the scalar field in question is determined by
prompt contribution and quasi-normal
ringing. Then, with the passage of time a definite oscillatory power-law
fall-off is manifest. We get the following power-law exponents $- 3.14$ and $-3.64$ for $d = 5,~6$, respectively.
Having in mind Eq.(\ref{osc1})
one can remark that we get good agreement between numerical calculations and analytically predicted values: $- 3.11$ and $- 3.61$, accordingly, that is the discrepancies are only
 $1\% $ and $0.8\% $. For all considered cases, 
the discrepancies between analytical and numerical values range from $0.2\% $ up to $1.8\% $. 
The period of oscillation is $T = \pi/ m \simeq 314.5 \pm 0.5$.
\par
Next, in Fig.2 we elaborate the evolution of the massive scalar field on the black hole future event horizon
$H_{+}$
approximated by $\psi(u = 10^{4},~t)$. We studied the evolution as a function of
$v$ for different spacetime dimensions, i.e., $d = 5,~6$. The calculations parameters were as follows:
$k = 0,~n = 1,~ B = 0.9,~ m = 0.01.$ The power-law exponents and the period of oscillation are nearly
the same as in Fig. \ref{fig1}.
Then, we studied the behaviour of $\psi$ due to the change of brane tension $B$. In our calculations we take
$B = 0.9,~0.7,~0.5$ (for $k = 0,~n = 1$), while $d = 5,~6$, respectively. The obtained results are presented in Figs.
 \ref{fig3} 
and \ref{fig4}. The received power-law exponents are $- 3.14,~ - 3.47,~- 4.04$ for the five-dimensional case and 
$- 3.64,~ - 3.97,~- 4.54$ for the six-dimensional
spacetime. We observe that the bigger $B$ is the smaller power-law exponents we get, i.e.,
the greater brane tension is exerted the faster decay of massive scalar hair on the considered
black hole spacetimes is observed.
\par
We also studied the power-law decay behaviour for different $n = 0,~1,~2$ (Fig.5) 
and established $B = 0.9$ and $k = 0$
in five-dimensional black hole spacetime, where we obtained the power-law exponents equal to 
$- 2.03,~ - 3.14,~- 4.26$. It turned out, that the same calculation parameters reveal in six-dimensional case
the power-law exponents of the form $- 2.53,~ - 3.64,~- 4.78$ (see Fig.6).
Figs. \ref{fig7} and \ref{fig8} depict the dependence of the decay pattern on the multipole number $k$.
The case of different masses of the scalar field in 
the black hole background was treated in Figs. \ref{fig9} and \ref{fig10}. Namely,
we studied the late-time decay rate on future timelike infinity in five-dimensional black hole spacetime for
different masses of the field in question (see Fig. \ref{fig9}). The decay rate was a power-law fall-off with the slope $-3.14$.
The six-dimensional case was presented in Fig. \ref{fig10}. For various masses of scalar field one gets the power-law
fall-off with the slope of the curve equal to $-3.64$.

\section{Conclusions}
\label{con}
Motivated by the forthcoming possibility of producing in LHC experiments microscopic black holes
and the necessity of studying their properties
we have elaborated the intermediate behaviour of massive scalar fields in the background of a static
$d$-dimensional tensional brane black hole. 
Due to the tremendous difficulties in solving equations of motion for the massive scalar field
in general $d$-dimensional black hole spacetime, we studied numerically the case of a five and six-dimensional
spacetime. The case of a six-dimensional black hole pierced by a topological defect is of a special interest
due to the interpretation of this line element as a black hole residing on a tensional 3-brane
embedded in a six-dimensional spacetime.
We checked analytical calculations by numerical ones and obtained good agreement with analytically predicted values.
Numerical studies of the intermediate late-time behaviour of massive scalar field on future timelike infinity
and on the future event horizon of tense brane black hole revealed the inverse power-law decay of the field in question.
The inverse power-law decay pattern was checked numerically for $d = 5,~6$. 
It turned out that the intermediate late time behaviour depended on the mass of the scalar field, as well as the multipole number of the wave mode and parameters characterizing topological defect.
The same situation takes place in the five-dimensional case. As we have mentioned, for a six-dimensional codimension-2 brane black hole
its event horizon radius is strictly connected with three-brane tension. Thus, our numerical
calculations revealed that the greater brane tension is the faster massive scalar field decays in the spacetime in question.

To finish with
let us give some remarks concerning the four-dimensional case. In four-dimensions the considered
line element reduce to the case found in \cite{ary86}. 
It describes a thin cosmic string passing through a  Schwarzschild black hole.
It was revealed that it constituted
the limit of a much more realistic situation when a black hole was pierced
by a Nielsen-Olesen vortex \cite{greg}. 
In the extremal limit of the black hole under consideration a completely
new phenomenon occurred. The flux of the vortex is expelled from the black hole
(for some range of black hole parameters), rather like the flux is expelled from
a superconductor (the so-called {\it Meissner effect}).
It happened that in the case of the extremal black holes
in dilaton gravity one has always expulsion of the Higgs fields from
their interiors \cite{mod99}. 
\par
Using the method presented in \cite{coe02}, it could be easily found that the eigenvalues
of the Laplace operator on a $S^2$ unit sphere pierced by a cosmic string, i.e.,
\mbox{$\la^2 = (k + \mid \tn \mid)~(k + \mid \tn \mid + 1)$},
where $k = 0, 1, 2 \dots$
It turned out that
the crucial role is played by the factor $B$ connected with a mass per unit length of a string, because of the fact that
$\tn = n / B$. Having in mind \cite{gibrog08,mod08} it can be shown that 
the intermediate asymptotic behaviour of a scalar 
field depends on the field's parameter mass, cosmic string parameter as well as the multipole number $k$.
The late-time behaviour reveals that the power law decay rate which is independent of 
the parameters in question has the form of $t^{-5/6}$.

%
%

\begin{acknowledgements}
This work was partially financed by the Polish budget funds in 2008 year as
the research project.
\end{acknowledgements}





\begin{acknowledgments}
This work was partially financed by the Polish budget funds in 2008 year as
the research project.
\end{acknowledgments}


\pagebreak

\pagebreak
\begin{figure}
\begin{center}
 \includegraphics[width=0.85\textwidth]{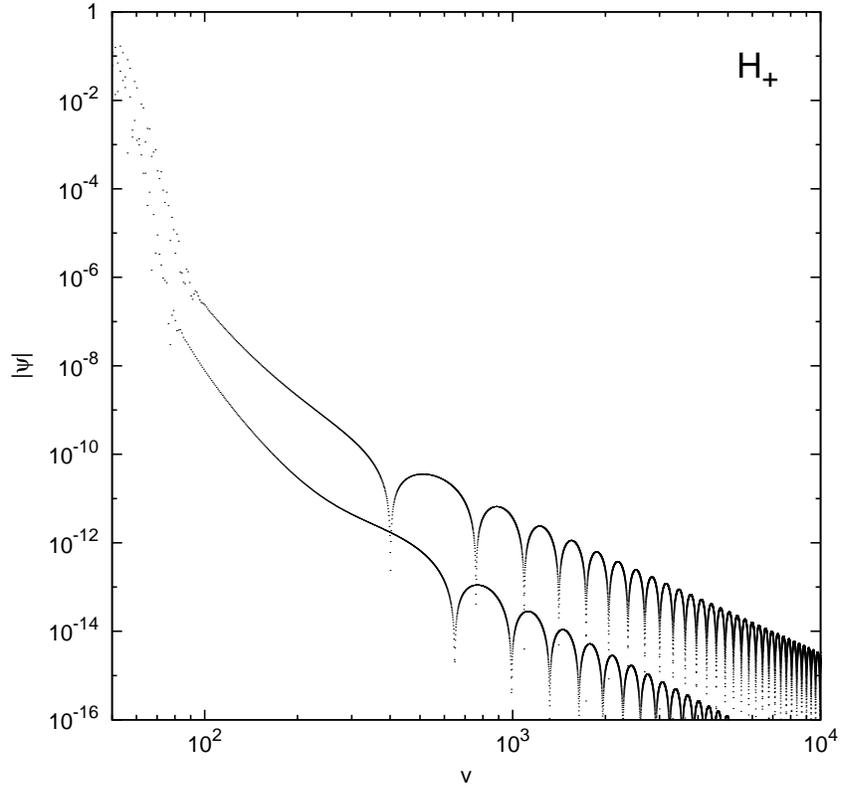}
\end{center}
\caption{Evolution of the scalar field $ \mid \psi \mid$ on the black hole future event horizon
$H_{+}$ (approximated by $\psi(u = 10^{4},~v)$) 
as a function of $v$ for different spacetime dimensions $d = 5,~6$, upper and lower curve, respectively. Calculation
parameters: 
$m = 0.01,~k = 0,~n = 1,~ B = 0.9$. The power-law exponents are:
$- 3.15$ and $- 3.67$, for the worst bottom curve.
The period of oscillation is 
\mbox{$T = {\pi \over m} \simeq 314.5 \pm 0.5$} to within
 $1 \% $ for both curves.}
\label{fig2}
\end{figure}

\pagebreak

\begin{figure}
\begin{center}
  \includegraphics[width=0.85\textwidth]{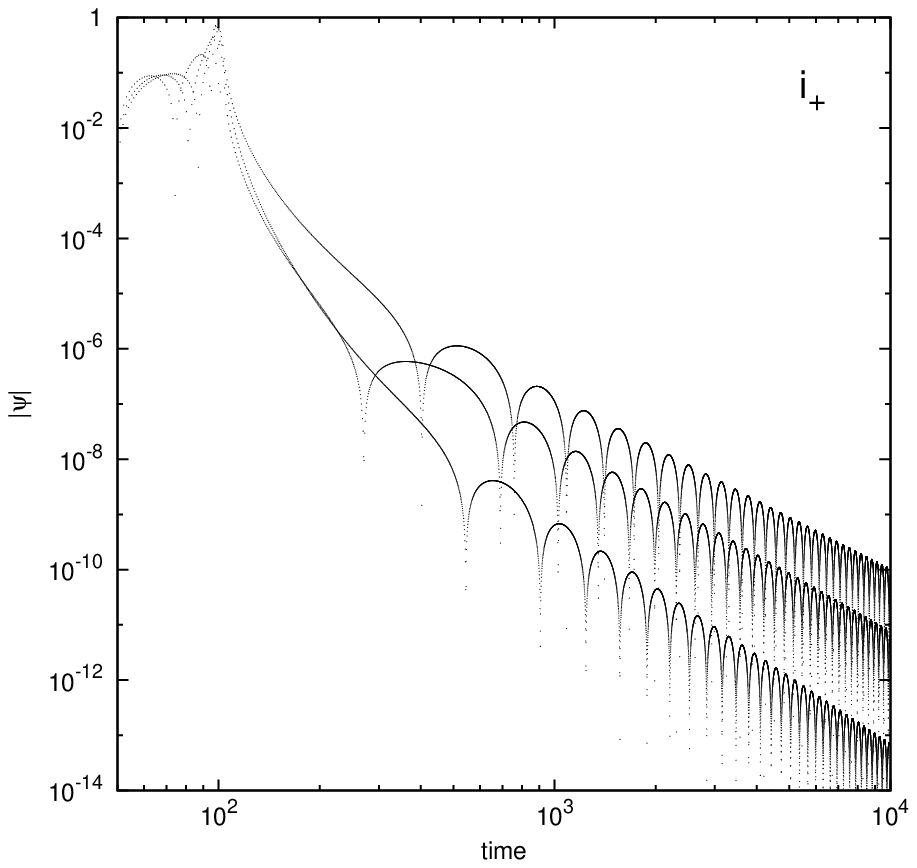}
\end{center}
\caption{Evolution of the scalar field $| \psi |$ on the future timelike infinity
$i_{+}$ as a function of $t$ for different \mbox{$B = 0.9$,} $0.7$, $0.5$ 
(curves from the top, respectively) and $d = 5$. Other calculation parameters: $m = 0.01$, \mbox{$k = 0$,}
 $n = 1$. The power-law exponents are:
$- 3.14,~- 3.47$ and $- 4.04$.
The period of oscillation is 
\mbox{$T = {\pi \over m} \simeq 314.5 \pm 0.5$ to within $0.2 \% $} for all curves.}
\label{fig3}
\end{figure}

\pagebreak

\begin{figure}
\begin{center}
  \includegraphics[width=0.85\textwidth]{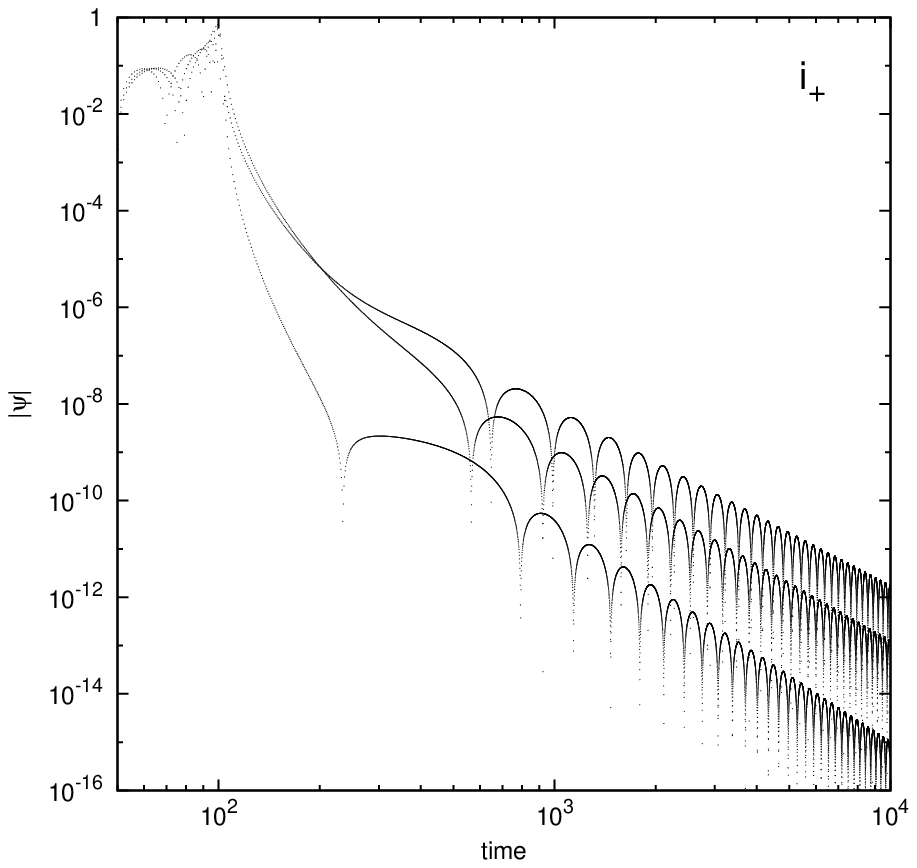}
\end{center}
\caption{Evolution of the scalar field $\mid \psi \mid$ on the future timelike infinity
$i_{+}$ as a function of $t$ for different $B = 0.9,~0.7,~0.5$ 
(curves from the top, respectively) in $d = 6$ spacetime. Other calculation parameters:
$m = 0.01,~k = 0,~n = 1$. The power-law exponents are:
$- 3.64,~- 3.97$ and $- 4.54$. The period of oscillation is $T = {\pi \over m} \simeq 314.5 \pm 0.5$ to within $0.4 \% $ for all curves.}
\label{fig4}
\end{figure}

\pagebreak
\begin{figure}
\begin{center}
  \includegraphics[width=0.85\textwidth]{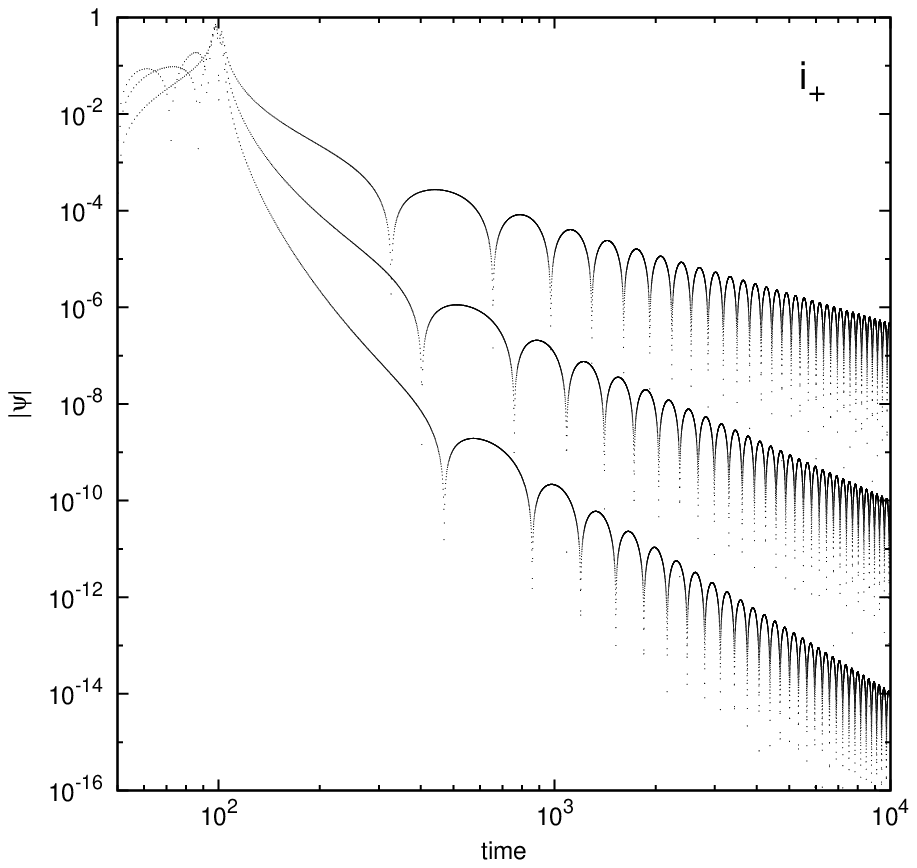}
\end{center}
\caption{Evolution of the scalar field $| \psi |$ on the future timelike infinity
$i_{+}$ as a function of $t$ for different $n = 0,~1,~2$ 
(curves from the top, respectively) in $d = 5$ spacetime. Other calculation parameters 
are:
\mbox{$m = 0.01,~k = 0,~n = 1$.} The power-law exponents are:
$- 2.03,~- 3.14$ and $- 4.26$. The period of oscillation is $T = {\pi \over m}
\simeq 314.5 \pm 0.5$ to within $0.3 \% $ for all curves.}
\label{fig5}
\end{figure}

\pagebreak
\begin{figure}
\begin{center}
  \includegraphics[width=0.85\textwidth]{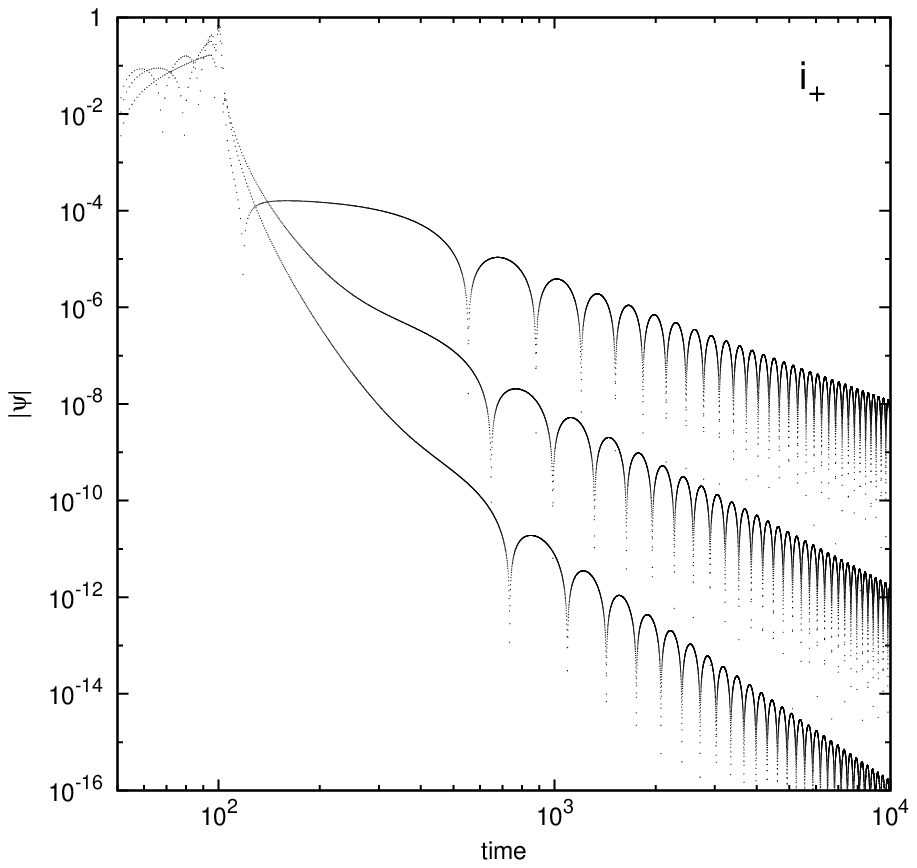}
\end{center}
\caption{Evolution of the scalar field $\mid \psi \mid$ on the future timelike infinity
$i_{+}$ as a function of $t$ for different $n = 0,~1,~2$ 
(curves from the top, respectively) in $d = 6$ spacetime. Other calculation parameters
in Fig.5.
The power-law exponents are:
$- 2.53,~- 3.64$ and $- 4.78$. The period of oscillation is $T = {\pi \over m}
\simeq 314.5 \pm 0.5$ to within $0.6 \% $ for all curves.}
\label{fig6}
\end{figure}

\pagebreak

\begin{figure}
\begin{center}
  \includegraphics[width=0.85\textwidth]{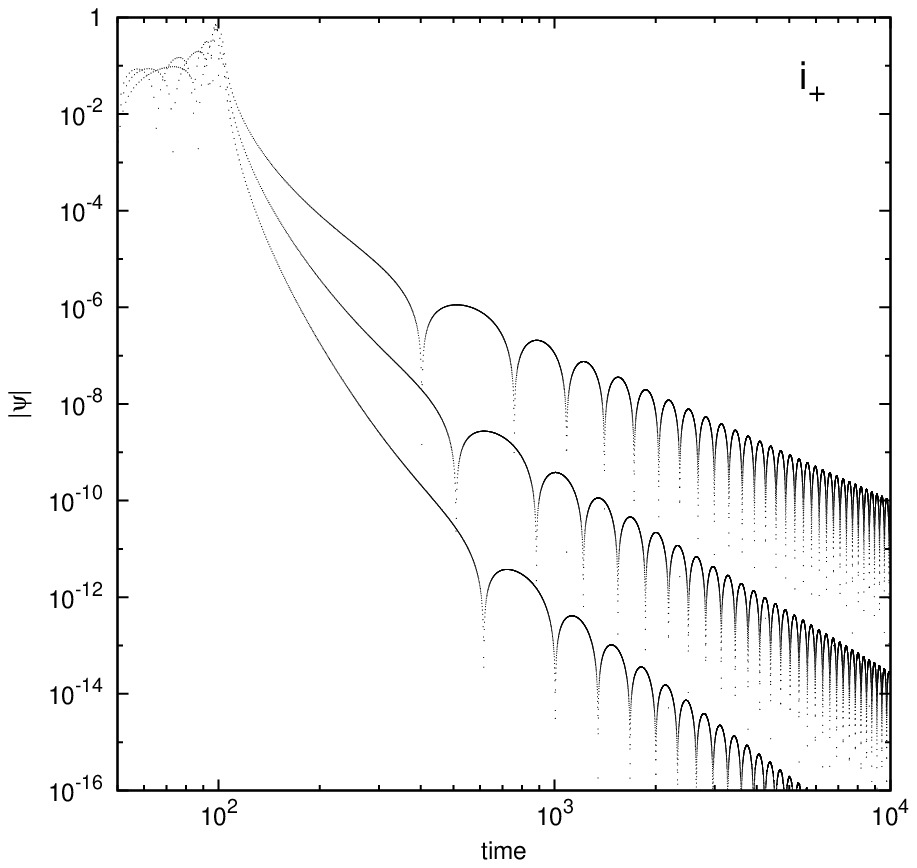}
\end{center}
\caption{Evolution of the scalar field $\mid \psi \mid$ on the future timelike infinity
$i_{+}$ as a function of $t$ for different $k = 0,~1,~2$ 
(curves from the top, respectively) in $d = 5$ spacetime. Other calculation parameters
are:
\mbox{$m = 0.01,~n = 1,~B = 0.9$.}
The power-law exponents are:
$- 3.14,~- 4.15$ and $- 5.20$. The period of oscillation is $T = {\pi \over m}
\simeq 314.5 \pm 0.5$ to within $1.7 \% $ for all curves.}
\label{fig7}
\end{figure}

\pagebreak
\begin{figure}
\begin{center}
  \includegraphics[width=0.85\textwidth]{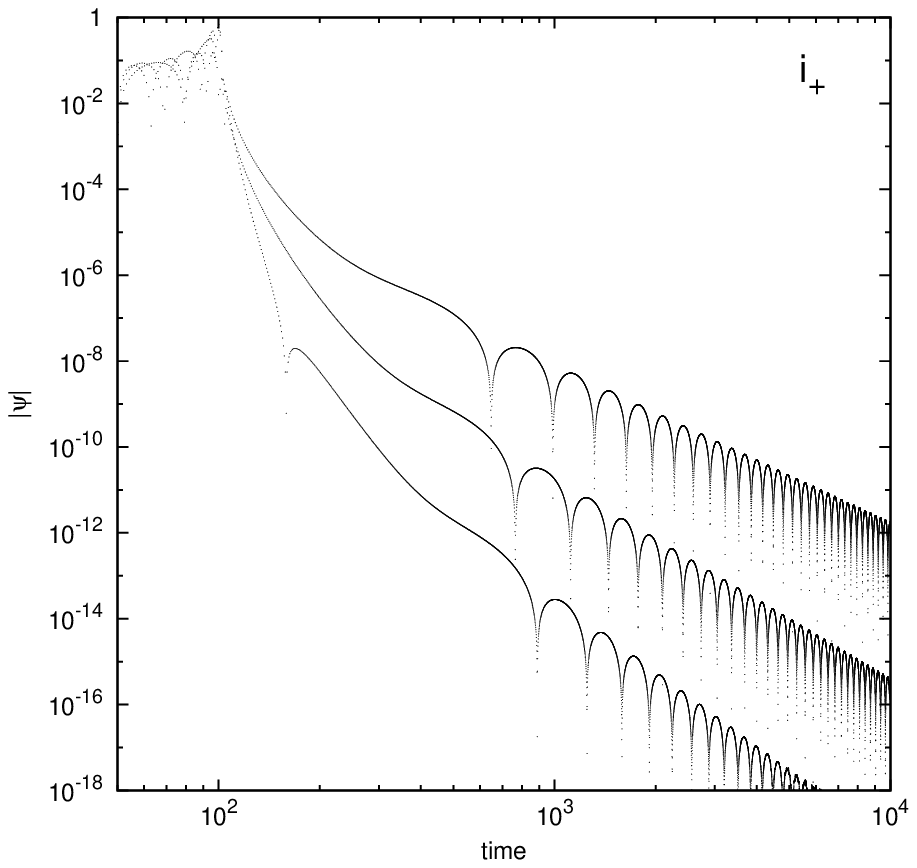}
\end{center}
\caption{Evolution of the scalar field $\mid \psi \mid$ on the future timelike infinity
$i_{+}$ as a function of $t$ for different $k = 0,~1,~2$ 
(curves from the top, respectively) in $d = 6$ spacetime. Other calculation parameters
in Fig.7.
The power-law exponents are:
$- 3.64,~- 4.67$ and $- 5.60$. The period of oscillation is $T = {\pi \over m}
\simeq 314.5 \pm 0.5$ to within $3.5 \% $ for all curves.}
\label{fig8}
\end{figure}

\pagebreak

\begin{figure}
\begin{center}
  \includegraphics[width=0.85\textwidth]{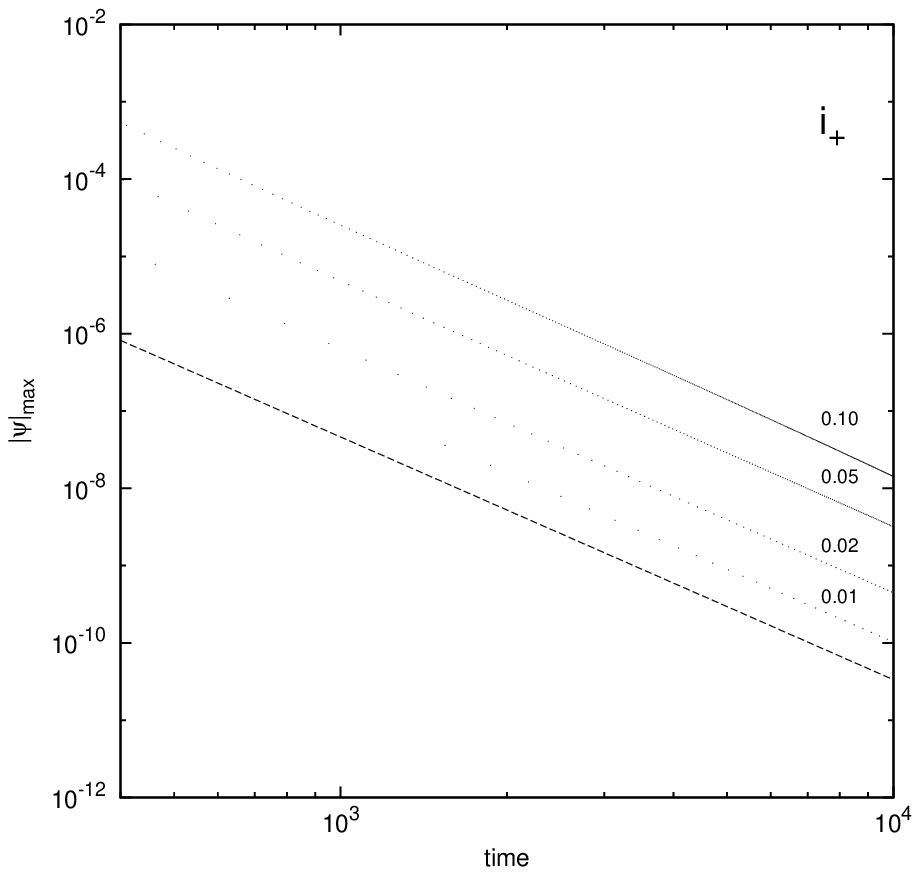}
\end{center}
\caption{Late-time decay rate of the field $\mid \psi \mid_{max}$ at $i_{+}$
for the five-dimensional case and different masses of the scalar field $m$ (values written above the lines).
Only maxima of the oscillations are showed. 
The dashed line has the slope equal to $- 3.14$.}
\label{fig9}
\end{figure}

\pagebreak

\begin{figure}
\begin{center}
  \includegraphics[width=0.85\textwidth]{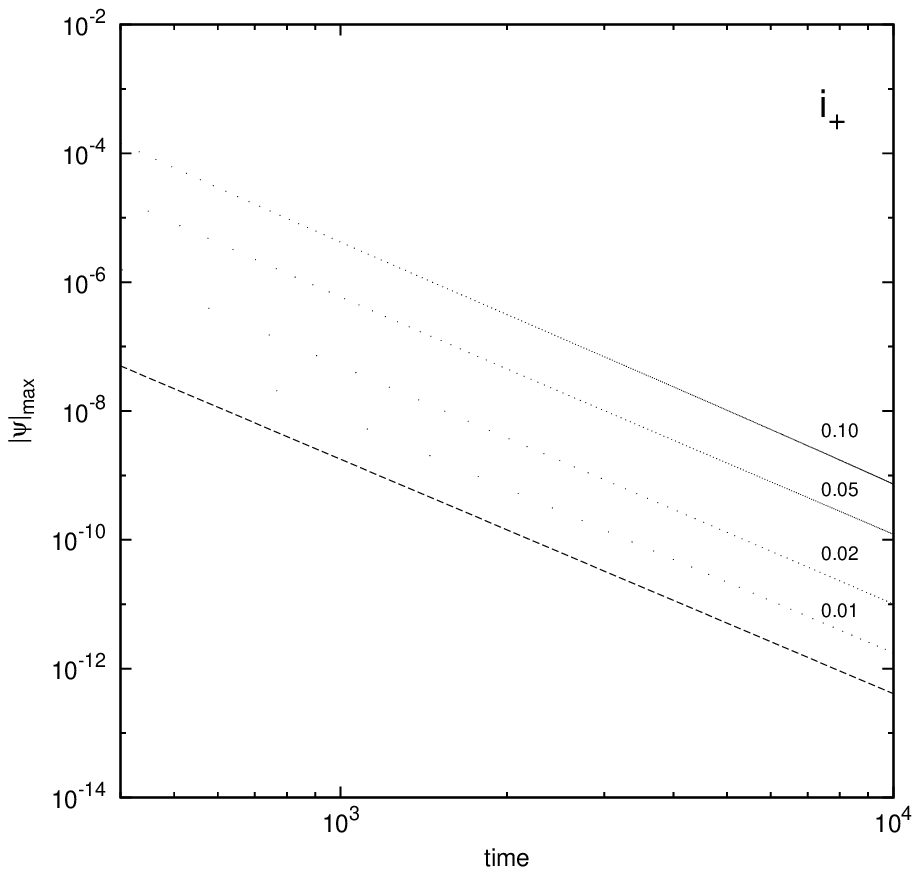}
\end{center}
\caption{Late-time decay rate of the field $\mid \psi \mid_{max}$ at $i_{+}$
for the six-dimensional case and different masses of the scalar field $m$ (values written above the lines).
Only maxima of the oscillations are showed. 
The dashed line has the slope equal to $- 3.64$.}
\label{fig10}
\end{figure}

\end{document}